\documentclass[prd,aps,showpacs,twocolumn]{revtex4}
\usepackage{graphicx}
\usepackage{bm}

\begin{document}

\title{Spin Hall effect of Photons in a Static Gravitational Field}
\author{Pierre Gosselin$^{1}$, Alain B\'{e}rard$^{2}$ and Herv\'{e} Mohrbach$^{2}$}

\address{$^1$ Universit\'e Grenoble I, Institut Fourier, UMR
5582 CNRS-UJF, UFR de Math\'ematiques, BP74, 38402 Saint Martin
d'H\`eres, Cedex, France \\
$^2$ Universit\'e Paul Verlaine, Institut de Physique, ICPMB1-FR
CNRS 2843, Laboratoire de Physique Mol\'eculaire et des
Collisions, 1, boulevard Arago, 57078 Metz, France}

\begin{abstract}
Starting from a Hamiltonian description of the photon within the set of
Bargmann-Wigner equations we derive new semiclassical equations of motion
for the photon propagating in static gravitational field. These equations
which are obtained in the representation diagonalizing the Hamiltonian at
the order $\hbar $, present the first order corrections to the geometrical
optics. The photon Hamiltonian shows a new kind of helicity-torsion
coupling. However, even for a torsionless space-time, photons do not follow
the usual null geodesic as a consequence of an anomalous velocity term. This
term is responsible for the gravitational birefringence phenomenon: photons
with distinct helicity follow different geodesics in a static gravitational
field.
\end{abstract}

\maketitle

\section{Introduction}

In the last few years many studies focused on the transport of quantum
particles with spin. Indeed, manipulating spin polarization of electrons is
a challenging goal in semiconductor spintronics. Achieving this goal
requires the understanding of the spin transport mechanism in systems with
spin orbit (SO) interaction. It was found that in such a system a Berry
phase in momentum space plays an important role by affecting both particle
phase and its transport properties \cite{ZHANG}. It is well known since the
seminal work of Berry \cite{BERRY}, that when a quantum mechanical system
has an adiabatic evolution, a wave function acquires a geometric phase. It
is only recently that the possible influence of the Berry phase on transport
properties (in particular on the semiclassical dynamics) of several physical
systems has been investigated. Semiconductors, having SO couplings greatly
enhanced with respect to the vacuum case, require a theory of spin
transport. However, even in the vacuum, new fundamental results concerning
the semiclassical equations of motion of electrons were recently derived.
For instance in \cite{ALAIN} and \cite{BLIOKH1}, considering the Dirac
equation in an external potential, it was shown that the position operator
acquires a spin-orbit contribution which turns out to be a Berry connection
rendering the coordinate algebraic structure non-commutative. This
drastically modifies the semiclassical equations of motion and implies a
topological spin transport similar to the intrinsic spin Hall effect in
semiconductors \cite{ZHANG}. A similar non-commutative algebra has been also
found in the context of electrons in magnetic Bloch bands \cite{NOUS},
leading to an anomalous velocity term.

Despite its very different nature, the photon displays many similar
behaviors with electronic phenomena such as energy bands in photonic
crystals and localization. These similarities stem from the wave-like nature
of quantum particles. Because a photon is also a spinning particle it is
important to understand if SO interaction may influence the transport of
light in a similar way as electrons in vacuum or in semiconductors. It has
been known for long that there is no position operator with commuting
components and as a consequence photons are not localizable. Therefore, one
of the main differences between photons and electrons also disappears as the
coordinates of both particles have in fact noncommuting components. It is
precisely this property which is at the origin of the SO interaction leaving
no doubt about its contribution to the propagation of photons. Therefore
these recent studies of the SO coupling in different systems taught us that
one can treat both kind of particles on an equal footing. It is thus even
more legitimate to wonder whether electronic phenomena have photonic
counterparts.

In addition, the localization of light rays is the essential ingredient of
the construction of Minkowski space-time. Also in the context of general
relativity, it has been argued that the Riemannian metric is determined by
the properties of light propagation \cite{EHLERS}\cite{HEHL}. A deeper
understanding of the properties of light and in particular its SO
interaction which is at the origin of the non localisability, is thus
necessary to build a quantum version of space-time.

It has already been observed that the SO coupling induces a rotation of the
polarization plane of light propagating in an optical fiber with torsion
\cite{CHIAO}. This effect had been predicted long ago by Rytov and
Vladimirskii \cite{RYTOV} \cite{VLADIMIRSKII} and can be interpreted in
terms of Berry phase \cite{BERRY} in momentum space. Recently it was shown
that the noncommutativity of the coordinates affects the ray of light itself
in an isotropic inhomogeneous medium \cite{ALAIN,BLIOKH2}. Other approaches
to the spin-orbit contributions for the propagation of light in isotropic
inhomogeneous media have been the focus of several other works \cite
{MURAKAMI} and have led to a generalization of geometric optics called
geometric spinoptics \cite{HORVATHY1}.

In this letter we investigate how the photon propagation in a static
gravitational field, which can be seen as an anisotropic inhomogeneous
medium, is affected by its spin-orbit interaction. Our photon description is
based on a Dirac like Hamiltonian in an arbitrary static gravitational field
with a possible (non-specified) torsion of space. The Bargmann-Wigner
equations allow us to build the wave function of a spin one particle and to
deduce the dynamical operators which satisfy unusual commutation relations
in the representation where the Hamiltonian has been diagonalized at the
order $\hbar $. It is in the diagonal representation (similarly to the Foldy
Wouthuysen (FW) representation) that the physical content of the theory is
best revealed. In particular we find that the photon semiclassical
Hamiltonian shows a kind of helicity-torsion coupling resulting from the
presence of Berry curvatures. From the semi-classical Hamiltonian we can
deduce new equations of motion taking Berry-curvatures into account. These
equations correspond to the first corrections to the geometrical optics. It
is found that the helicity is always conserved and that even in a
torsionless space-time, photons do not follow the null geodesic due to an
anomalous velocity term. The presence of this term is a general feature of
the modern point of view of the topological spin transport of quantum
particles \cite{ZHANG}\cite{ALAIN}\cite{BLIOKH1}\cite{BLIOKHAP}. The
anomalous velocity is directly responsible for the gravitational
birefringence phenomenon, i.e., photons with distinct helicity follow
different geodesics in a static gravitational field. This confirms earlier
claims suggesting that the absence of birefringence in Einstein's gravity
might be only a consequence of the limitation of the geometrical optics
limit \cite{MASHOON}. Our computation also shows that the velocity of light
is still equal to $c$ at this order of approximation (a polarization
dependent velocity would be a $\hbar ^{2}$ phenomenon). Finally, it is worth
noticing that there exists another spinorial representation of the photon
Hamiltonian equivalent to the Maxwell equation \cite{ILLGE} which contrary
to the Bargmann-Wigner equations does not suffer inconsistencies due to the
redundancy of the photon wave functions. However it can be seen that both
approaches lead to the same semi-classical energy and dynamical operators.
The advantage of the Bargmann-Wigner approach presented here is that it is
more usual, more straightforward and that it removes also the null energy
eigenstates.

\section{Free photon}

We start with the description of the photon dynamics in the vacuum by
considering the two Bargmann-Wigner equations ($r=1,2$) of a massless spin $%
1 $ particle \cite{BARGMANN}:
\begin{equation}
\left( \gamma _{0}^{(r)}\partial _{0}+\gamma _{i}^{(r)}\partial ^{i}\right)
\Psi _{(a_{1}a_{2})}(x)=0\ \ \ \ \ \ \ (i=1,2,3)  \label{bargman}
\end{equation}
where$\Psi _{\left( a_{1}a_{2}\right) }=\Psi _{(a_{2}a_{1})}$ is the
symmetrized Bargmann-Wigner amplitude with $a_{i\text{ }}$running from $1$
to $4$ the spinorial indices of the wave function (as usual the indices $0$
and $i$ refer respectively to the local time and space coordinates, with the
flat metric ($+,-,-,-$) ). The $\gamma ^{(r)}$ matrix is a $\gamma $ matrix
acting on $a_{i}$. This symmetrized direct product of two Dirac spinors
assures that the positive energy subspace forms a $3$-d space corresponding
to the irreducible representation of angular momenta $1$ deduced from the
composition of two states with angular momentum $1/2$. It can be proved that
Eqs.\ref{bargman} are equivalent to the Maxwell equations \cite{LIVRE}.

Next, we write an Hamiltonian associated to each Bargmann-Wigner equations $%
\hat{H}^{(r)}=\mathbf{\alpha }^{(r)}\mathbf{.P}$ which can be diagonalized
by the product of usual FW unitary transformations \cite{FW}, $U^{(r)}(%
\mathbf{P})=(E+c\beta \mathbf{\alpha }^{(r)}\mathbf{.P})/E\sqrt{2}$ such
that
\begin{equation}
U^{(1)}(\mathbf{P})U^{(2)}(\mathbf{P})\hat{H}^{(r)}U^{(1)}(\mathbf{P}%
)^{+}U^{(2)}(\mathbf{P})^{+}=E\beta ^{(r)},
\end{equation}
with $E=Pc$ the energy of the photon in the vacuum. The wave function is
also transformed and becomes $\phi _{(a_{1}a_{2})}(\mathbf{P}%
)=E^{-1}(U^{(1)}(\mathbf{P})^{+}U^{(2)}(\mathbf{P})^{+})\Psi _{(a_{1}a_{2})}(%
\mathbf{P})$ in the FW representation. Whereas the quasi-momentum is
invariant through the action of $U^{(r)}$, i.e., $\widetilde{\mathbf{p}}%
\mathbf{=}U^{(1)}U^{(2)}\mathbf{P}U^{(1)+}U^{(2)+}=\mathbf{P}$, the position
operator becomes $\widetilde{\mathbf{r}}=i\hbar \partial _{\mathbf{p}}+%
\mathcal{A}^{(1)}+\mathcal{A}^{(2)}$ with a pure gauge potential $\mathcal{A}%
^{(r)}=i\hbar U^{(r)}(\mathbf{p})\partial _{\mathbf{p}}U^{(r)}(\mathbf{p}%
)^{+}$ induced by the FW transformation. After projection of $\widetilde{%
\mathbf{r}}$ on the positive energy subspace we get the non trivial $SU(2)$
connection $\mathbf{A=}(\mathbf{p}\wedge \mathbf{\Sigma })/p^{2}$, and a new
position operator $\mathbf{r}$ for the free photon $\mathbf{r}=\mathbf{R}+(%
\mathbf{p}\wedge \mathbf{\Sigma })/p^{2}$ where $\mathbf{R=}i\hbar \partial
_{\mathbf{p}}$ is the canonical coordinate operator and $\mathbf{\Sigma }%
=\hbar (\mathbf{\sigma }^{(1)}+\mathbf{\sigma }^{(2)})/2$ the spin one
matrix. This definition of the position operator gives rise, through the
commutation relations, to a monopole in momentum space $[x^{i},x^{j}]=i\hbar
\theta ^{ij}(\mathbf{p})=-i\hbar \varepsilon ^{ijk}\lambda p_{k}/p^{3}$
where $\lambda =\pm \hbar $ is the helicity and $\theta ^{ij}(\mathbf{p}%
)=\nabla _{p^{i}}A_{j}-\nabla _{p^{j}}A_{i}+[A_{i,}A_{j}]$ the so called
Berry curvature (the origin of this name is explained in \cite{ALAIN}, see
also \cite{ALAIN2}).

We then build the Hamiltonian of the free photon of positive energy as $%
H=(E^{(1)}+E^{(2)})/2=pc$ with $E^{(r)}=pc$ the positive eigenvalue of the
operator $\hat{H}^{(r)}$. Therefore, for free particles, the dynamical
equations of motion in the FW representation are trivially given by $d%
\mathbf{r}/dt=\frac{i}{\hbar }[\mathbf{r},H]=\mathbf{p}c/p$ and $d\mathbf{p}%
/dt=\frac{i}{\hbar }[\mathbf{p},H]=0$ which in particular implies for the
light velocity $v=c$. Note that the same results can be deduced from the
Maxwell equations \cite{VIETELLO}.

\section{Photon in a static gravitational field.}

We now extend our previous approach to the case of a photon propagating in a
arbitrary static gravitational field, where $g_{0i}=0$ for $i=1,2,3$, so
that $ds^{2}=g_{00}(dx^{0})^{2}-g_{ij}dx^{i}dx^{j}=0$. Consider again the
Bargmann-Wigner equations of motion with the following associated
Hamiltonian of the Dirac form
\begin{equation}
\hat{H}^{(r)}=\sqrt{g_{00}}\mathbf{\alpha }^{(r)}.\mathbf{\tilde{P}}+\frac{%
\hbar }{4}\varepsilon _{\varrho \beta \gamma }\Gamma _{0}^{\varrho \beta
}\sigma ^{\gamma }+i\frac{\hbar }{4}\Gamma _{0}^{0\beta }\alpha _{\beta }
\label{HGR}
\end{equation}
and $\mathbf{\tilde{P}}$ given by $\tilde{P}_{\alpha }\mathbf{=}h_{\alpha
}^{i}(\mathbf{R})(P_{i}+\frac{\hbar }{4}\varepsilon _{\varrho \beta \gamma
}\Gamma _{i}^{\varrho \beta }\sigma ^{\gamma })$ with $h_{\alpha }^{i}$ the
static orthonormal dreibein $(\alpha =1,2,3)$, $\Gamma _{i}^{\alpha \beta }$
the spin connection components and $\varepsilon _{\alpha \beta \gamma
}\sigma ^{\gamma }=\frac{i}{8}(\gamma ^{\alpha }\gamma ^{\beta }-\gamma
^{\beta }\gamma ^{\alpha }).$ The coordinate operator is again given by $%
\mathbf{R=}i\hbar \partial _{\mathbf{p}}.$ Note that here we consider the
general case where an arbitrary static torsion of space is allowed. It is
known \cite{LECLERC} that for a static gravitational field (which is the
case considered here), the Hamiltonian $\hat{H}^{(r)}$ is Hermitian. We now
want to diagonalize $\hat{H}^{(r)}$ through a unitary transformation $%
U^{(r)}(\mathbf{\tilde{P}}).$ Because the components of $\mathbf{\tilde{P}}$
depend both on operators $\mathbf{P}$ and $\mathbf{R}$ the diagonalization
at order $\hbar $ is performed by adapting the method detailed in \cite
{METHODE} to block-diagonal Hamiltonians. To do so, we first write $\hat{H}%
^{(r)}$ in a symmetrical way in $\mathbf{P}$ and $\mathbf{R}$ at first order
in $\hbar $. This is easily achieved using the Hermiticity of the
Hamiltonian which yields \
\[
\hat{H}^{(r)}=\frac{1}{2}\left( \sqrt{g_{00}}\mathbf{\alpha }^{(r)}.\mathbf{%
\tilde{P}+\tilde{P}}^{+}\mathbf{.\alpha }^{(r)}\sqrt{g_{00}}\right) +\frac{%
\hbar }{4}\varepsilon _{\varrho \beta \gamma }\Gamma _{0}^{\varrho \beta
}\sigma ^{\gamma }.
\]
Now for a ease of exposition we temporarily include the $g_{00}$ is the
definition of the vierbein $h_{\alpha }^{i}\rightarrow g_{00}h_{\alpha
}^{i}. $ The contributions of the $g_{00}$ will be again explicitly written
later on in the final result for the energy.

\subsection{Semiclassical Hamiltonian Diagonalization}

The semiclassical diagonalization is achieved in two steps. In the first one
we diagonalize at first order in $\hbar $ by considering the formal
situation where $\mathbf{R}$ is considered as a parameter commuting with $%
\mathbf{P}$. In the second one we add the contributions due to the non
commuting character of the dynamical variables.

\subsubsection{Diagonalization when $P$ and $R$ commute.}

The Hamiltonian $\hat{H}_{0}^{(r)}$ (we add the index $0$ when $\mathbf{R}$
is a parameter) can then be diagonalized at first order in $\hbar $ by the
following unitary FW matrix
\begin{eqnarray}
U_{0}^{(r)}(\mathbf{\tilde{P})} &=&\frac{D}{\sqrt{2(E_{0}^{^{(r)}})^{2}}}%
\left( E_{0}^{(r)}+c\frac{1}{2}\beta \left( \mathbf{\alpha }^{(r)}.\mathbf{%
\tilde{P}}\right. \right.  \nonumber \\
&&\left. \left. \mathbf{+\tilde{P}}^{+}\mathbf{.\alpha }^{(r)}\right)
+N\right)
\end{eqnarray}
with $E_{0}^{(r)}=\sqrt{\left( \frac{\mathbf{\alpha }^{(r)}.\mathbf{\tilde{P}%
+\tilde{P}}^{+}\mathbf{.\alpha }^{(r)}g}{2}\right) ^{2}}$. We introduced
also the notations $N=\frac{\hbar }{4}\frac{i\mathbf{\alpha }^{(r)}.\left(
\mathbf{P\times \Gamma }_{0}\right) }{P}$ and $D=1+\frac{\hbar }{4}\beta
\frac{\left( \mathbf{P\times \Gamma }_{0}\right) \times \mathbf{P}}{2P^{3}}$
with $\Gamma _{0\gamma }=\varepsilon _{\varrho \beta \gamma }\Gamma
_{0}^{\varrho \beta }$. (here, and in the sequel, $P$ means $\left(
P^{2}\right) ^{\frac{1}{2}}$, with the metric $g_{ij}$ and $P^{3}$ is
obviously defined as $\left( P^{2}\right) ^{\frac{3}{2}}$).

The proof of this diagonalization relies on the following properties: For
each parameter $\mathbf{R}$ the matrices $h_{\alpha }^{i}$ and $\Gamma
_{i}^{\alpha \beta }$ are independent of both the momentum and position
operators. The matrices $\beta $ and $\mathbf{\alpha .\tilde{P}}$
anticommute and in the Taylor expansion of $E_{0}^{(r)}$ all terms commute
with $\beta $ and $\mathbf{\alpha }^{(r)}\mathbf{.\tilde{P}+\mathbf{\tilde{P}%
}}^{+}\mathbf{\mathbf{.\alpha }^{(r)}}$. In this context the diagonalized
Hamiltonian is equal to $U_{0}^{(r)}\hat{H}_{0}^{(r)}U_{0}^{(r)+}$\ which
reads
\begin{eqnarray}
U_{0}^{(r)}\hat{H}_{0}^{(r)}U_{0}^{(r)+} &=&c\beta ^{(r)}\sqrt{\left( \frac{%
\mathbf{\alpha }^{(r)}.\mathbf{\tilde{P}+\tilde{P}}^{+}\mathbf{.\alpha }%
^{(r)}}{2}\right) ^{2}}  \nonumber \\
&&+\frac{\hbar \left( \mathbf{P.\Gamma }_{0}\right) \mathbf{P.\sigma }^{(r)}%
}{4P^{2}}  \label{EO}
\end{eqnarray}
which can also be written
\begin{eqnarray}
U_{0}^{(r)}\hat{H}_{0}^{(r)}U_{0}^{(r)+} &=&c\beta ^{(r)}\sqrt{\mathbf{P}%
^{2}+\frac{\hbar }{2}\mathbf{P.\sigma }^{(r)}\varepsilon _{\varrho \beta
\gamma }\Gamma _{i}^{\varrho \beta }h^{i\gamma }}  \nonumber \\
&&+\frac{\hbar \left( \mathbf{P.\Gamma }_{0}\right) \mathbf{P.\sigma }^{(r)}%
}{4P^{2}}
\end{eqnarray}
In this last expression we have neglected contributions of curvature type of
order $\hbar ^{2}$.

\subsubsection{Corrections when $\mathbf{P}$ and $\mathbf{R}$ do not commute}

Now, to diagonalize $H^{(r)}$ at the semi-classical approximation, it is
shown in \cite{METHODE} that it is enough to apply the following FW
transformation ($\mathbf{R}$ being now an operator)
\begin{equation}
U^{(r)}(\mathbf{\tilde{P})}=U_{0}^{(r)}(\mathbf{\tilde{P})}+X^{(r)}
\label{FW2}
\end{equation}
with
\begin{equation}
X^{(r)}=\frac{i}{4\hbar }\left[ \mathcal{A}_{P^{l}}^{(r)},\mathcal{A}%
_{R_{l}}^{(r)}\right] U^{(r)}(\mathbf{\tilde{P})}
\end{equation}
where we defined the position and momentum pure gauge Berry potential $%
\mathcal{A}_{\mathbf{R}}^{(r)}=i\hbar U^{(r)}\nabla _{\mathbf{P}}U^{(r)+}$
and $\mathcal{A}_{\mathbf{P}}^{(r)}=-i\hbar U^{(r)}\nabla _{\mathbf{R}%
}U^{(r)+},$ and then to project the transformed Hamiltonian $U^{(r)}\hat{H}%
^{(r)}U^{(r)+}$ on the positive energy states. All expressions in $U^{(r)}(%
\mathbf{\tilde{P})}$ are implicitly assumed to be symmetrized in $\mathbf{P}$
and $\mathbf{R}$ and the corrective term $X^{(r)}$ must be added to restore
the unitarity of $U^{(r)}(\mathbf{\tilde{P})}$ which is destroyed by the
symmetrization. After projection on the positive energy subspace needed to
perform the diagonalization \cite{METHODE} the resulting position and
momentum operators can thus be written $\mathbf{r}^{\left( r\right) }=%
\mathbf{R}+\mathbf{A}_{R}^{(r)}$ and $\mathbf{p}^{\left( r\right) }=\mathbf{P%
}+\mathbf{A}_{P}^{(r)}$ where the explicit computation gives for the
components
\begin{equation}
A_{P,k}^{(r)}=-\hbar c^{2}\frac{\varepsilon ^{\alpha \beta \gamma }\tilde{P}%
_{\alpha }\mathbf{\sigma }_{\beta }^{(r)}(\mathbf{\nabla }_{R_{k}}\tilde{P}%
_{\gamma })}{2E^{(r)2}}+O(\hbar ^{2})
\end{equation}
\begin{equation}
A_{R_{k}}^{(r)}=\hbar c^{2}\frac{\varepsilon ^{\alpha \beta \gamma
}h_{\gamma }^{k}\tilde{P}_{\alpha }\mathbf{\sigma }_{\beta }^{(r)}}{2E^{(r)2}%
}+O(\hbar ^{2})
\end{equation}
with $E^{(r)}$ the same as $E_{0}^{(r)}$ above but now $\mathbf{R}$ is an
operator. Performing our diagonalization process leads us to the following
expression for the positive energy operator
\begin{eqnarray}
\tilde{\varepsilon}^{(r)} &\simeq &c\sqrt{p^{2}+\frac{\hbar }{2}\mathbf{%
p.\sigma }^{(r)}\varepsilon _{\varrho \beta \gamma }\Gamma _{i}^{\varrho
\beta }h^{i\gamma }}+\frac{\hbar }{4}\frac{\left( \mathbf{p.\Gamma }%
_{0}\right) \mathbf{p.\sigma }^{(r)}}{p^{2}}  \nonumber \\
&&-\frac{i\hbar }{4\varepsilon ^{(r)}}\left[ \nabla _{R_{l}}(\mathbf{\alpha }%
^{(r)}\mathbf{.\tilde{P})},\nabla _{P^{l}}(\mathbf{\alpha }^{(r)}\mathbf{.%
\tilde{P})}\right]   \nonumber \\
&&+(\nabla _{R_{l}}\varepsilon ^{(r)})A_{R_{l}}^{(r)}+(\nabla
_{P^{l}}\varepsilon ^{(r)})A_{P^{l}}^{(r)}+O(\hbar ^{2})  \label{EEE}
\end{eqnarray}
Some computations allow us to rewrite the r.h.s. of Eq. \ref{EEE} in a more
familiar form.\ Define first the $\gamma $-component of the vector $\mathbf{%
\Gamma }_{i}$ as $\Gamma _{i,\gamma }=\varepsilon _{\varrho \beta \gamma
}\Gamma _{i}^{\varrho \beta }(\mathbf{r})$ and the helicity $\lambda ^{(r)}=%
\frac{\hbar \mathbf{p.\sigma }^{(r)}}{p}.$ Introducing also
\[
\varepsilon ^{(r)}=c\sqrt{\left( p_{i}+\frac{\lambda ^{(r)}}{4}\frac{\Gamma
_{i}(\mathbf{r}).\mathbf{p}}{p}\right) g^{ij}\left( p_{j}+\frac{\lambda
^{(r)}}{4}\frac{\Gamma _{j}(\mathbf{r}).\mathbf{p}}{p}\right) },
\]
an explicit computation shows that the semiclassical energy reads
\[
\tilde{\varepsilon}^{(r)}=\varepsilon ^{(r)}+\frac{\lambda ^{(r)}}{4}\frac{%
\mathbf{p.\Gamma }_{0}}{p}+\frac{\hbar \mathbf{B}.\mathbf{\sigma }^{(r)}}{%
2\varepsilon ^{(r)}}-\frac{(\mathbf{A}_{R}^{(r)}\mathbf{\times p).B}}{%
\varepsilon ^{(r)}}.
\]
where we have introduced a field $B_{\gamma }=-\frac{1}{2}P_{\delta
}T^{\alpha \beta \delta }\varepsilon _{\alpha \beta \gamma }$ with $%
T^{\alpha \beta \delta }=h_{k}^{\delta }\left( h^{l\alpha }\partial
_{l}h^{k\beta }-h^{l\beta }\partial _{l}h^{k\alpha }\right) +h^{l\alpha
}\Gamma _{l}^{\beta \delta }-h^{l\beta }\Gamma _{l}^{\alpha \delta }$ the
usual torsion for a static metric (where only space indices in the
summations give non zero contributions).

Interestingly, this semi-classical Hamiltonian presents formally the same
form as the one of a Dirac particle in a true external magnetic field \cite
{BLIOKH1}\cite{METHODE}. The term $\mathbf{B}.\mathbf{\sigma }$ is
responsible for the Stern-Gerlach effect, and the operator $\mathbf{L}=(%
\mathbf{A}_{R}\mathbf{\times p)}$ is the intrinsic angular momentum of
semiclassical particles. The same contribution appears also in the context
of the semiclassical behavior of Bloch electrons (spinless) in an external
magnetic field \cite{NOUS}\cite{NIU} where it corresponds to a magnetization
term. Because of this analogy and since $T^{\alpha \beta \delta }$ is
directly related to the torsion of space through $T^{\alpha \beta \delta
}=h_{k}^{\delta }h^{i\alpha }h^{j\beta }T_{ij}^{k}$ we call $\mathbf{B}$ a
magnetotorsion field.

However, this form for the energy presents the default to involve the spin
rather than the helicity, this last quantity being more fundamental for a
photon. Actually one can use the property $\lambda \mathbf{p}/2p=\hbar
\mathbf{\sigma /}2-(\mathbf{A}_{R}\mathbf{\times p)}$ to rewrite the energy
as
\begin{equation}
\tilde{\varepsilon}^{(r)}=\varepsilon ^{(r)}+\frac{\lambda ^{(r)}}{4}\frac{%
\mathbf{p.\Gamma }_{0}}{p}+\frac{\lambda ^{(r)}}{2\varepsilon ^{(r)}}\frac{%
\mathbf{B}.\mathbf{p}}{p}.
\end{equation}
We can now build the Hamiltonian as the sum of the two Hamiltonians for
one-half massless spinning particle $\widetilde{\varepsilon }=(\tilde{%
\varepsilon}^{(1)}+\tilde{\varepsilon}^{(2)})/2$. By Taylor expanding the
expression of $\varepsilon ^{(r)}$ we see that at the leading order in $%
\hbar $ the sum does recombine to give after reintroducing the $g_{00}$
dependence
\begin{equation}
\widetilde{\varepsilon }\simeq \varepsilon +\frac{\lambda }{4}\frac{\mathbf{%
p.\Gamma }_{0}}{p}+\frac{\lambda g_{00}}{2\varepsilon }\frac{\mathbf{B}.%
\mathbf{p}}{p}  \label{NUMBER}
\end{equation}
where $\varepsilon =c\sqrt{\left( p_{i}+\frac{\lambda }{4}\frac{\Gamma _{i}(%
\mathbf{r}).\mathbf{p}}{p}\right) g^{ij}g_{00}\left( p_{j}+\frac{\lambda }{4}%
\frac{\Gamma _{j}(\mathbf{r}).\mathbf{p}}{p}\right) }$ with $\lambda
=(\lambda ^{(1)}+\lambda ^{(2)})/2$ the photon helicity. In the same manner,
the Berry connections for the photon become $\mathbf{A}_{R}=\mathbf{A}%
_{R}^{(1)}+\mathbf{A}_{R}^{(2)}$ and $\mathbf{A}_{P}=\mathbf{A}_{P}^{(1)}+%
\mathbf{A}_{P}^{(2)}$. This allow us to write the dynamical operators at
leading order in $\hbar $ as
\begin{eqnarray}
\mathbf{r} &=&i\hbar \partial _{\mathbf{p}}+\hbar c^{2}\frac{\mathbf{P}%
\times \mathbf{\Sigma }}{2\varepsilon ^{2}}  \label{rGB} \\
\mathbf{p} &=&\mathbf{P}-\hbar c^{2}(\frac{\mathbf{P}\times \mathbf{\Sigma }%
}{2\varepsilon ^{2}})\mathbf{\nabla }_{\mathbf{R}}\mathbf{\tilde{P}}
\label{pGB}
\end{eqnarray}
where in Eqs\ref{rGB} and \ref{pGB} $\varepsilon ^{2}$ can be approximated
by $c^{2}p_{i}g^{ij}g_{00}p_{j}$ at the order considered.

The semi-classical Hamiltonian Eq.\ref{NUMBER} is one of the main results of
this paper. It contains, in addition to the energy term $\varepsilon $, new
contributions due to the Berry connections. Indeed, Eq.\ref{NUMBER} shows
that the helicity couples to the gravitational field through the
magnetotorsion field $\mathbf{B}$ which is non-zero for a space with
torsion. As a consequence, a hypothetical torsion of space may be revealed
through the presence of this coupling. Note that, in agrement with \cite
{SILENKO}, this Hamiltonian does not contain the spin-gravity coupling term $%
\mathbf{\Sigma .\nabla }g_{00}$ predicted in \cite{OBUKOV}.

From Eqs.\ref{rGB} and \ref{pGB} we deduce the new (non-canonical)
commutations rules
\begin{eqnarray}
\left[ r^{i},r^{j}\right] &=&i\hbar \Theta _{rr}^{ij} \\
\left[ p^{i},p^{j}\right] &=&i\hbar \Theta _{pp}^{ij} \\
\left[ p^{i},r^{j}\right] &=&-i\hbar g^{ij}+i\hbar \Theta _{pr}^{ij}
\end{eqnarray}
where $\Theta _{\zeta \eta }^{ij}=\partial _{\zeta ^{i}}A_{\eta
^{j}}-\partial _{\eta ^{i}}A_{\zeta ^{j}}+[A_{\zeta ^{i},}A_{\eta ^{j}}]$
where $\zeta ,\eta $.mean either $r$ or $p$. An explicit computation shows
that at leading order
\begin{eqnarray}
\Theta _{rr}^{ij} &=&-\hbar c^{4}\frac{\left( \mathbf{\Sigma }.\mathbf{p}%
\right) p_{\gamma }}{2\varepsilon ^{4}}\varepsilon ^{\alpha \beta \gamma
}h_{\alpha }^{i}h_{\beta }^{j}  \nonumber \\
\Theta _{pp}^{ij} &=&-\hbar c^{4}\frac{\left( \mathbf{\Sigma }.\mathbf{p}%
\right) p_{\gamma }}{2\varepsilon ^{4}}\nabla _{r_{i}}p_{\alpha }\nabla
_{r_{j}}p_{\beta }\varepsilon ^{\alpha \beta \gamma }  \nonumber \\
\Theta _{pr}^{ij} &=&\hbar c^{4}\frac{\left( \mathbf{\Sigma }.\mathbf{p}%
\right) p_{\gamma }}{2\varepsilon ^{4}}\nabla _{r_{i}}p_{\alpha }h_{\beta
}^{j}\varepsilon ^{\alpha \beta \gamma }
\end{eqnarray}
From the additional commutation relations between the helicity and the
dynamical operators $[r_{i},\lambda ]=[p_{i},\lambda ]=0$ we deduce the
semiclassical equations of motion
\begin{eqnarray}
\mathbf{\dot{r}} &=&\left( 1-\Theta _{pr}\right) \nabla _{\mathbf{p}}\tilde{%
\varepsilon}+\mathbf{\dot{p}\times }\Theta _{rr}  \nonumber \\
\mathbf{\dot{p}} &=&-\left( 1-\Theta _{pr}\right) \nabla _{\mathbf{r}}\tilde{%
\varepsilon}+\mathbf{\dot{r}}\times \Theta _{pp}  \label{Eqmotion}
\end{eqnarray}
To complete the dynamical description of the photon notice that at the
leading order the helicity $\lambda $ is not changed by the unitary
transformation which diagonalizes the Hamiltonian so that it can be written $%
\lambda =\hbar \mathbf{p}.\mathbf{\Sigma }/p$. After a short computation one
can check that the helicity is always conserved
\begin{equation}
\frac{d}{dt}\left( \frac{\hbar \mathbf{p.\Sigma }}{p}\right) =0
\end{equation}
for an arbitrary static gravitational field independently of the existence
of a torsion of space.

Eqs.\ref{Eqmotion} are the new semiclassical equations of motion for a
photon in a static gravitational field. They describe the ray trajectory of
light in the first approximation of geometrical optics (GO). (In GO it is
common to work with dimensionless momentum operator $\mathbf{p=}k_{0}^{-1}%
\mathbf{k}$ with $k_{0}=\omega /c$ instead of the momentum \cite{BLIOKH2}).
For zero Berry curvatures we obtain the well known zero order approximation
of GO and photons follow the null geodesic. The velocity equation contains
the by now well known anomalous contribution $\mathbf{\dot{p}\times }\Theta
^{rr}$ which is at the origin of the intrinsic spin Hall effect (or Magnus
effect) of the photon in an isotropic inhomogeneous medium of refractive
index $n(r)$ \cite{ALAIN,BLIOKH2,MURAKAMI,HORVATHY1}. Indeed, this term
causes an additional displacement of photons of distinct helicity in
opposite directions orthogonally to the ray. Consequently, we predict
gravitational birefringence since photons with distinct helicities follow
different geodesics. In comparison to the usual velocity $\mathbf{\dot{r}}%
=\nabla _{\mathbf{p}}\tilde{\varepsilon}$ $\sim c$, the anomalous velocity
term $\mathbf{v}_{\perp }$ is obviously small, its order $v_{\perp }^{i}\sim
c\widetilde{\lambda }\nabla _{r^{j}}g^{ij}$ being proportional to the wave
length $\widetilde{\lambda }$.

The momentum equation presents the dual expression $\mathbf{\dot{r}}\times
\Theta _{pp}$ of the anomalous velocity which is a kind of Lorentz force
which being of order $\hbar $ does not influence the velocity equation at
order $\hbar $. Note that similar equations of motion with dual
contributions $\mathbf{\dot{p}\times }\Theta _{rr}$ and $\mathbf{\dot{r}}%
\times \Theta _{pp}$ were predicted for the wave-packets dynamics of
spinless electrons in crystals subject to small perturbations \cite{NIU}.
The complicated Eq. \ref{Eqmotion} simplifies greatly for a symmetric
gravitational field as shown below.

\section{Symmetric gravitational field}

As a simple application, consider the symmetric case $g_{00}g^{ij}=\delta
^{ij}F^{2}(\mathbf{R})$. A typical example of such a metric is the
Schwarzschild space-time in isotropic coordinates. For a symmetric metric
one has $\mathbf{B.p=\Gamma }_{0}=0$ and the semiclassical energy Eq. \ref
{NUMBER} reduces to
\begin{equation}
\widetilde{\varepsilon }=c(pF(\mathbf{r})+F(\mathbf{r})p)/2
\end{equation}
with the dynamical variables $\mathbf{r}=\mathbf{R}+\hbar \frac{\mathbf{P}%
\times \mathbf{\Sigma }}{P^{2}}$, $\mathbf{p}=\mathbf{P}$, and the following
commutation relations $[r^{i},r^{j}]=i\hbar \Theta _{rr}^{ij}=-i\hbar
\lambda \varepsilon ^{ijk}p_{k}/p^{3}$, $[p^{i},p^{j}]=0$, $%
[r^{i},p^{j}]=i\hbar g^{ij}$. As a consequence, we derive the following
equations of motion
\begin{eqnarray}
\mathbf{\dot{r}} &=&\nabla _{\mathbf{p}}\tilde{\varepsilon}+\mathbf{\dot{p}%
\times }\Theta _{rr}  \nonumber \\
\mathbf{\dot{p}} &=&-\nabla _{\mathbf{r}}\tilde{\varepsilon}
\end{eqnarray}
In the symmetric case the equations of motion become simpler than in the
general case, but the gravitational birefringence is still present. These
equations were already postulated (but not derived) in \cite{ALAIN} to
explain the Magnus effect (the different deviation of light of distinct
polarization in an inhomogeneous medium of refractive index $n(r)$) observed
despite its smallness in inhomogeneous isotropic optical fibers \cite
{ZELDOVITCH} and also discussed theoretically in less general contexts and
with different approaches in several other papers \cite{BLIOKH2} \cite
{MURAKAMI}\cite{HORVATHY1}. This case fits within our formalism since a
gravitational field can be seen as an \ isotropic medium related to the
metric through the relation $g^{ij}=\delta ^{ij}n^{-1}(r)$. Therefore the
gravitational birefringence predicted here is simply due to the Magnus
effect as a consequence of the photon spin-orbit interaction. In particular
this effect does not need a coupling between the electromagnetic field and a
torsion term as proposed in \cite{RUBILAR}.

We now apply the equations of motion to compute the deflection of polarized
light by a star's gravitational field. A polarization independent result is
expected by the Einstein's theory of gravitation which does not consider the
anomalous velocity. With the Schwarzschild metric one has $F(R)=1-\frac{2GM}{%
R}$ \cite{OBUKOV} and for the equations of motion we get
\begin{eqnarray*}
\stackrel{\cdot }{\mathbf{p}} &=&-2GM\frac{\mathbf{r}}{r^{3}}p \\
\mathbf{\dot{r}} &=&\frac{\mathbf{p}}{p}F+\lambda \frac{2GM}{r^{3}}\frac{%
\mathbf{r\times p}}{p^{2}}
\end{eqnarray*}
We can therefore easily compute the angle of deflection $\Delta \phi $
between in-and outgoing polarized rays
\[
\Delta \phi =\frac{4GM}{c^{2}r_{0}}\left( 1-\frac{\lambda }{\hbar }\frac{%
\widetilde{\lambda }}{2\pi r_{0}}\right)
\]
where $r_{0}$ is the smallest distance of the light ray to the central
source of gravitation, $M$ the mass of the star, $\widetilde{\lambda }$ the
wave length of the photon and $\lambda $ the helicity.

We observe that the deviation is helicity dependent as a consequence of the
anomalous velocity, but that the effect is very small being of order $%
\widetilde{\lambda }/r_{0}$ and certainly unobservable for a star like the
Sun.

\section{Conclusion}

In this paper, we have diagonalized at the first order in $\hbar $ the
photon Hamiltonian in a static gravitational field. This diagonal
Hamiltonian displays a new interaction between the helicity and a torsion
field. As a consequence the torsion of space, if any, could in principle be
determined through this coupling. However, even in the absence of torsion,
we found two new semiclassical equations of motion including Berry phase
contributions for both dynamical variables, predicting that the photon does
not follow the null geodesic due to its spinning nature. The reason is an
anomalous velocity, responsible for the gravitational birefringence. This
last result is in agreement with the modern point of view about the spinning
particles evolution. Our results are not restricted to the gravitational
field but also apply to systems with anisotropic refractive indices.

\end{document}